\documentstyle[amsmath,amssymb,epsfig,12pt]{article}

\begin{document}
\setlength{\parskip}{0.45cm}
\setlength{\baselineskip}{0.75cm}
%
%
%
\begin{titlepage}
\setlength{\parskip}{0.25cm}
\setlength{\baselineskip}{0.25cm}
\begin{flushright}
DO-TH 06/08\\
\vspace{0.2cm}
DESY 06-178\\
\vspace{0.2cm}
September 2006
\end{flushright}
\vspace{1.0cm}
\begin{center}
\Large
{\bf The curvature of $F_2^p(x,Q^2)$ as a probe of the range of}
\\\Large{\bf validity of perturbative QCD evolutions in the small-$x$ region}
\vspace{1.5cm}

\large
M.~Gl\"uck$^a$, C.~Pisano$^b$, E.\ Reya$^a$\\
\vspace{1.0cm}

\normalsize
$^a${\it Universit\"{a}t Dortmund, Institut f\"{u}r Physik}\\
{\it D-44221 Dortmund, Germany} \\
\vspace{0.5cm}
$^b${\it Universit\"{a}t Hamburg, II.~Institut f\"{u}r Theoretische Physik}\\
{\it Luruper Chaussee 149, D-22761 Hamburg, Germany}

\vspace{1.5cm}
\end{center}

\begin{abstract}
\noindent Perturbative NLO and NNLO QCD evolutions of parton distributions
are studied, in particular in the (very) small-$x$ region, where they are 
in very good agreement with all recent precision measurements of 
$F_2^p(x,Q^2)$.  These predictions turn out to be also rather insensitive
to the specific choice of the factorization scheme ($\overline{\rm MS}$ or
DIS).  A characteristic feature of perturbative QCD evolutions is a 
{\em{positive}} curvature of $F_2^p$ which increases as $x$ decreases.
This perturbatively stable prediction provides a sensitive test of the 
range of validity of perturbative QCD.
\end{abstract}
\end{titlepage}


\section{Introduction}
The curvature of DIS structure functions like $F_2^p(x,Q^2)$, i.e., its
second derivative with respect to the photon's virtuality $Q^2$ at
fixed values of $x$, plays a decisive role in probing the range of 
validity of perturbative QCD evolutions of parton distributions in the 
small-$x$ region.  This has been observed recently \cite{ref1,ref2} and
it was demonstrated that NLO($\overline{\rm MS}$) evolutions
imply a {\em positive} curvature which increases as $x$ decreases.
However, in contrast to \cite{ref1} where this positive curvature
was shown to disagree with the data, the conventional {\em full} NLO
analysis performed in \cite{ref2} led to the conclusion that no such
disagreement prevails. It was therefore concluded \cite{ref2} that 
the NLO small-$x$ parton evolution equations are {\em not} challenged
by the small-$x$ data on $F_2^p$.
These rather unique predictions provide a check of the range of validity
of perturbative QCD evolutions.  However, the curvature is a rather 
subtle mathematical quantity which a priori may sensitively depend on
the theoretical (non)perturbative assumptions made for calculating it.
The main purpose of the present article is to study the dependence and
stability of the predicted curvature with respect to a different choice
of the factorization scheme (DIS versus $\overline{\rm MS}$) and to 
the perturbative order of the evolutions by extending the common NLO
(2-loop) evolution \cite{ref2} to the next-to-next-to-leading 3-loop
order (NNLO).

\section{Theoretical formalism}

In the common $\overline{\rm MS}$ factorization scheme the relevant
$F_2^p$ structure function as extracted from the DIS $ep$ process can be,
up to NNLO, written as \cite{ref3,ref4,ref5}
\begin{equation}
F_2^p(x,Q^2) = F_{2,{\rm NS}}^+(x,Q^2)+ F_{2,S}(x,Q^2)+F_2^c(x,Q^2,m_c^2)
\end{equation}
with the non--singlet contribution for three active (light) flavors 
being given by 
\begin{equation}
\frac{1}{x}\, F_{2,{\rm NS}}^+(x,Q^2) =
\Big[C_{2,q}^{(0)}+aC_{2,{\rm NS}}^{(1)}+a^2C_{2,{\rm NS}}^{(2)+} 
\Big] 
\otimes 
\left[ \frac{1}{18}\, q_8^+ +\frac{1}{6}\, q_3^+\right](x,Q^2)
\end{equation}
where $a=a(Q^2)\equiv\alpha_s(Q^2)/4\pi$, $C_{2,q}^{(0)}(z)=\delta(1-z)$,
$C_{2,{\rm NS}}^{(1)}$ is the common NLO coefficient function (see,
for example, \cite{ref6}) and a convenient expression for the relevant 
NNLO 2-loop Wilson coefficient $C_{2,{\rm NS}}^{(2)+}$ can be found 
in \cite{ref3}.  The NNLO $Q^2$-evolution of the flavor 
{\mbox{non-singlet} 
combinations $q_3^+=u+\bar{u}-(d+\bar{d})=u_v-d_v$ and $q_8^+=
u+\bar{u}+d+\bar{d}-2(s+\bar{s}) = u_v+d_v+4\bar{q}-4\bar{s}$, where
$\bar{q}\equiv\bar{u}=\bar{d}$ and $s=\bar{s}$, is related to the 
3-loop splitting function \cite{ref7} $P_{\rm NS}^{(2)+}$, besides 
the usual LO (1-loop) and NLO (2-loop) ones, $P_{\rm NS}^{(0)}$ and
$P_{\rm NS}^{(1)+}$, respectively \cite{ref3,ref8}.  Notice that we 
do not consider sea breaking effects ($\bar{u}\neq\bar{d},\,\, s\neq
\bar{s}$) since the HERA data used, and thus our analysis, are not
sensitive to such corrections.  The flavor singlet contribution in
(1) reads 
\begin{equation}
\frac{1}{x}\, F_{2,S}(x,Q^2)=\frac{2}{9}
\left\{ \left[ C_{2,q}^{(0)} + aC_{2,q}^{(1)}+a^2C_{2,q}^{(2)}\right]
\otimes \Sigma +
\left[aC_{2,g}^{(1)}+a^2C_{2,g}^{(2)}\right] \otimes g\right\} (x,Q^2)
\end{equation}
with 
$\Sigma(x,Q^2)\equiv\Sigma_{q=u,d,s}(q+\bar{q})=u_v+d_v+4\bar{q}+2\bar{s}$,
$C_{2,q}^{(1)}=C_{2,\rm NS}^{(1)}$ and the additional common NLO gluonic
coefficient function $C_{2,g}^{(1)}$ can be again found in \cite{ref6},
for example.  Convenient expressions for the NNLO $C_{2,q}^{(2)}$ and
$C_{2,g}^{(2)}$ have been given in \cite{ref4} and the relevant 3-loop
splitting functions $P_{ij}^{(2)}$, required for the evolution of 
$\Sigma(x,Q^2)$ and $g(x,Q^2)$, have been derived in \cite{ref9}.  We
have performed all $Q^2$-evolutions in Mellin $n$-moment space and used
the QCD-PEGASUS program \cite{ref10} for the NNLO evolutions.  In NNLO
the strong coupling evolves according to 
$da/d\ln Q^2 = -\Sigma_{\ell =0}^2\, \beta_{\ell}\, a^{\ell +2}$ where
$\beta_0 = 11-2f/3$, $\beta_1=102-38f/3$ and $\beta_2=2857/2-5033f/18+
325f^2/54$ and the running $a(Q^2)$ is appropriately matched at 
$Q=m_c=1.4$ GeV and $Q=m_b=4.5$ GeV. The heavy flavor
(charm) contribution $F_2^c$ in (1) is taken as in \cite{ref2} as given
by the fixed-order NLO perturbation theory \cite{ref11}.  The small 
bottom contribution turns out to be negligible for our purposes.  Notice
that a NNLO calculation of heavy quark production is not yet available.
For definiteness we work in the fixed flavor factorization scheme,
given in (1)-(3), rather than in the variable (massless quark) scheme
since the results for $F_2^p$ and its curvature remain essentially 
unchanged \cite{ref2}.

The choice of a factorization scheme in NLO, other than the
$\overline{\rm MS}$ scheme used thus far, might imply similar effects
as the additional NNLO contributions in the $\overline{\rm MS}$
scheme.  For example, in the deep inelastic scattering (DIS)
factorization scheme \cite{ref12,ref5,ref6} the Wilson coefficients
in (2) and (3) are absorbed into the parton distributions, or more
precisely into their evolutions, i.e., into the splitting functions.
Disregarding for simplicity all NNLO contributions, this transformation
to the DIS scheme in NLO is achieved via \cite{ref4,ref5}
\begin{equation}
P_{\rm NS}^{(1)}\to P_{\rm NS,\,DIS}^{(1)} = P_{\rm NS}^{(1)}
   +\beta_0\Delta C_{2,\rm NS}^{(1)}
\end{equation}
\begin{equation}
\hat{P}^{(1)}\to\hat{P}_{\rm DIS}^{(1)} = \hat{P}^{(1)} +
 \beta_0\Delta \hat{C}_2^{(1)} -
\left[\Delta \hat{C}_2^{(1)}\otimes \hat{P}^{(0)}-\hat{P}^{(0)}
 \otimes \Delta\hat{C}_2^{(1)}\right]
\end{equation}
where
\begin{equation}
\Delta C_{2,\rm NS}^{(1)} = - C_{2,\rm NS}^{(1)}\quad, \quad\quad
\Delta \hat{C}_2^{(1)} = -
\left( 
\renewcommand{\arraystretch}{1.5}\begin{array}{ccc}
C_{2,q}^{(1)} & , &  C_{2,g}^{(1)}\\
-C_{2,q}^{(1)} & , & -C_{2,g}^{(1)} \end{array} \right)\, .
\end{equation}
Instead of (2) and (3), the light $u,\, d,\, s$ quark contributions 
to $F_2^p$ in the NLO(DIS) factorization scheme now simply become
\begin{eqnarray}
F_2^p(x,Q^2) & = & x\sum_{q=u,d,s} e_q^2
   \left[q(x,Q^2)+\bar{q}(x,Q^2)\right]_{\rm DIS}+F_2^c\nonumber\\
& = & x
   \left[\frac{1}{18}\, q_8^+(x,Q^2)+\frac{1}{6}\, 
           q_3^+(x,Q^2)\right]_{\rm DIS}
   +\frac{2}{9}x\Sigma(x,Q^2)_{\rm DIS} +F_2^c\,\, .
\end{eqnarray}
The quantitative difference between the NLO($\overline{\rm MS}$)
and NLO(DIS) results will turn out to be rather small.  Therefore
we do not consider any further the DIS scheme in NNLO. 

Having obtained the parton distributions 
$\stackrel{(-)}{q}\!\!(x,Q^2)_{\rm DIS}$ and $g(x,Q^2)_{\rm DIS}$
from an explicit NLO analysis of $F_2(x,Q^2)$ in the DIS factorization
scheme, one can transform them to the $\overline{\rm MS}$ scheme
via (see \cite{ref13}, for example)
\begin{eqnarray}
\stackrel{(-)}{q}\!\!(x,Q^2) & = & \stackrel{(-)}{q}\!\!(x,Q^2)_{\rm DIS}
 -a\left[C_{2,q}^{(1)}\otimes \stackrel{(-)}{q}\!\!_{\rm DIS} +
   \frac{1}{2f}\, C_{2,g}^{(1)}\otimes g_{\rm DIS}\right]\!
     (x,Q^2)+{\cal{O}}(a^2)\\
g(x,Q^2) & = & g(x,Q^2)_{\rm DIS}
 +a\left[ C_{2,q}^{(1)}\otimes \Sigma_{\rm DIS} + C_{2,g}^{(1)}
    \otimes g_{\rm DIS}\right]\!
      (x,Q^2) +{\cal{O}}(a^2)
\end{eqnarray}
where
\begin{eqnarray}
C_{2,q}^{(1)}(z)&  = & 2\frac{4}{3}
 \left[\frac{1+z^2}{1-z} \left(\ln \frac{1-z}{z} -\frac{3}{4}\right)
  +\frac{1}{4}\, (9+5z)\right]_+\\
C_{2,g}^{(1)}(z) & = & 4f\frac{1}{2} 
 \left[(z^2+(1-z)^2)\ln\frac{1-z}{z} -1+8z(1-z)\right]
\end{eqnarray}
with $f=3$.  This transformation to the $\overline{\rm MS}$ scheme
then allows for a consistent comparison of our NLO(DIS) results
with the higher-order results obtained in the $\overline{\rm MS}$
factorization scheme.

\section{Quantitative results}
For the present analysis the valence $q_v=u_v,\, d_v$ and sea
$w=\bar{q},\, g$ distributions are parametrized at an input scale
$Q_0^2=1.5$ GeV$^2$ as follows:
\begin{eqnarray}
x\,q_v(x,Q_0^2) & = & N_{q_v}x^{a_{q_v}}(1-x)^{b_{q_v}}
  (1+c_{q_v}\sqrt{x}+d_{q_v}x + e_{q_v}x^{1.5})\\
x\,w(x,Q_0^2) & = & N_w x^{a_w}(1-x)^{b_w}(1+c_w\sqrt{x}+d_w x)
\end{eqnarray}
and without loss of generality the strange sea is taken to be 
$s=\bar{s}=0.5\, \bar{q}$.  The normalizations $N_{u_v}$ and $N_{d_v}$
are fixed by $\int_0^1 u_v dx = 2$ and $\int_0^1 d_v dx=1$,
respectively, and $N_g$ is fixed via $\int_0^1 x(\Sigma +g)dx=1$.
We have somewhat extended the set of DIS data used in \cite{ref2}
in order to determine the remaining parameters at larger values
of $x$ and of the valence distributions.  The following data sets
have been used:  the small-$x$ \cite{ref14} and large-$x$ 
\cite{ref15} H1 $F_2^p$ data; the fixed target BCDMS data 
\cite{ref16} for $F_2^p$ and $F_2^n$ using $Q^2\geq 20$ GeV$^2$
and $W^2=Q^2(\frac{1}{x}-1)+m_p^2\geq 10$ GeV$^2$ cuts, and the 
proton and deuteron NMC data \cite{ref17} for $Q^2\geq 4$ GeV$^2$
and $W^2\geq 10$ GeV$^2$.  This amounts to a total of 740 data 
points.  The required overall normalization factor of the data
turned out to be 0.98 for BCDMS and 1.0 for NMC.
The resulting parameters of the various fits are summarized in
Table 1.  The relevant small-$x$ predictions are compared with
the H1 data \cite{ref14} in Fig.~1, which are also consistent with
the ZEUS data \cite{ref18} with partly lower statistics.  The
present more detailed NLO($\overline{\rm MS}$) analysis corresponds
to $\chi^2/{\rm dof}=715.3/720$ and the results are comparable
to our previous ones \cite{ref2}. 
Our new NLO(DIS) and NNLO(3-loop) fits are also very similar, 
corresponding to $\chi^2/{\rm dof}=714.2/720$ and $712.0/720$, 
respectively, although they fall slightly
below the common NLO($\overline{\rm MS}$) predictions at smaller
values of $Q^2$. It should be emphasized that the perturbatively
stable QCD predictions are in perfect agreement with all recent
high-statistics measurements of the $Q^2$-dependence of 
$F_2^p(x,Q^2)$ in the (very) small-$x$ region.  Therefore
additional model assumptions concerning further resummations of
subleading small-$x$ logarithms (see, for example, \cite{ref19})
are not required \cite{ref7,ref9}.

In Figs.~2 and 3 we show our gluon and sea
input distributions in (13) and Table 1 as obtained in our three
different fits, as well as their evolved shapes at $Q^2=4.5$ GeV$^2$
in particular in the small-$x$ region. In order to allow for a
consistent comparison in the $\overline{\rm MS}$ scheme, our
NLO(DIS) results have been transformed to the $\overline{\rm MS}$
factorization scheme using (8) and (9).  Note, however, that the
gluon distribution in the DIS scheme is very similar to the one
obtained in NLO($\overline{\rm MS}$) shown in Fig.~2 which holds in
particular in the small-$x$ region.  This agreement becomes even
better for increasing values of $Q^2$.  This agreement is similar
for the sea distributions in the small-$x$ region shown in Fig.~3.
Only for 
$x$ \raisebox{-0.1cm}{$\stackrel{>}{\sim}$} 0.1 the NLO(DIS) sea
density becomes sizeably smaller than the NLO($\overline{\rm MS}$)
one shown in Fig.~3. The NLO results are rather similar but
distinctively different from the NNLO ones in the very small-$x$
region at $Q^2>Q_0^2$.  In particular the strong increase of the
gluon distribution $xg(x,Q^2)$ as $x\to 0$ at NLO is somewhat
tamed by NNLO 3-loop effects \cite{ref9}. 

Turning now to the curvature of $F_2^p$ we first present in Fig.~4
our results for $F_2^p(x,Q^2)$ at $x=10^{-4}$, together with a 
global fit MRST01 NLO result \cite{ref20}, as a function of \cite{ref1}
\begin{equation}
q = \log_{10}\left(1+\frac{Q^2}{0.5\,\,{\rm GeV}^2}\right)\, \, .
\end{equation}
This variable has the advantage that most measurements lie along
a straight line \cite{ref1} as indicated by the dotted line in 
Fig.~4.  All our three NLO and NNLO fits give almost the same
results which are also very similar \cite{ref2} to the global
CTEQ6M NLO fit \cite{ref21}.  In contrast to all other fits shown in
Fig.~4, only the MRST01 parametrization results in a sizeable
curvature for $F_2^p$ \cite{ref2}.  More explicitly the curvature
can be directly extracted from
\begin{equation}
F_2^p(x,Q^2) = a_0(x) + a_1(x)q + a_2(x)q^2\,\, .
\end{equation}
The curvature $a_2(x)=\frac{1}{2}\,\partial_q^2\,  F_2^p(x,Q^2)$ is
evaluated by fitting this expression to the predictions for
$F_2^p(x,Q^2)$ at fixed values of $x$ to a (kinematically) given
interval of $q$.  In Fig.~5a we present $a_2(x)$ which results 
from experimentally selected $q$-intervals \cite{ref1,ref2}:
\begin{eqnarray}
0.7 \leq q \leq 1.4\quad\quad & {\rm for} & \quad\quad
    2\times 10^{-4} < x < 10^{-2}\nonumber\\
0.7 \leq q \leq 1.2\quad\quad & {\rm for} & \quad\quad
    5\times 10^{-5} < x \leq 2\times 10^{-4}\, .
\end{eqnarray}
It should be noticed that the average value of $q$ decreases with
decreasing $x$ due to the kinematically more restricted $Q^2$ range
accessible experimentally. (We deliberately do not show the results at
the smallest available $x=5\times 10^{-5}$ where the $q$-interval is
too small, $0.6\leq q\leq 0.8$, for fixing $a_2(x)$ in (15) uniquely
and where moreover present measurements are not yet sufficiently
accurate \cite{ref1,ref2}). For comparison we also show in 
Fig.~5b the curvature $a_2(x)$ for an $x$-independent fixed $q$-interval
\begin{equation}
0.6\leq q \leq 1.4 \quad\quad\quad 
      (1.5\leq Q^2\leq 12\,\, {\rm GeV}^2)\, \, .
\end{equation}  
Apart from the rather large values of $a_2(x)$ specific \cite{ref2}
for the MRST01 fit, our NLO and NNLO results agree well with the 
experimental curvatures as calculated and presented in \cite{ref1}
using the H1 data \cite{ref14}.  Our predictions do 
{\em{not}} sensitively depend on the factorization scheme
chosen ($\overline{\rm MS}$ or DIS) and are, moreover, perturbative
{\em{stable}} with the NNLO 3-loop results lying typically
below the NLO ones, i.e.\ closer to present data.  It should be
emphasized that the perturbative stable evolutions always result
in a {\em{positive}} curvature which {\em{increases}}
as $x$ decreases.  Such unique predictions provide a sensitive
test of the range of validity of perturbative QCD!  
This feature is supported by the data shown in Fig.~5a. Future
analyses of present precision measurements in this very small-$x$
region (typically
$10^{-5}$ \raisebox{-0.1cm}{$\stackrel{<}{\sim}$} $x$
\raisebox{-0.1cm}{$\stackrel{<}{\sim}$} $10^{-3}$) should 
provide additional tests of the theoretical predictions concerning
the range of validity of perturbative QCD evolutions.

Finally, the question arises whether the second derivative of 
$F_2^p$ with respect to the variable $q$ in (15) is indeed dominated
by the curvature $\ddot{F}_2^p\equiv \partial^2F_2^p/\partial(\ln Q^2)^2$
which is directly related to the evolution equations and to experiment,
since $\partial_q^2F_2^p\equiv \partial^2 F_2^p/\partial q^2$ is a
linear combination of $\dot{F}_2^p\equiv\partial F_2^p/\partial \ln Q^2=
{\cal{O}}(\alpha_s)$ and $\ddot{F}_2^p={\cal{O}}(\alpha_s^2)$:
\begin{equation}
\partial_q^2 F_2^p = \left(
 \frac{Q^2+0.5\, {\rm GeV}^2}{Q^2}\, \ln 10\right)^2
 \left[-\kappa\, \dot{F}_2^p +\ddot{F}_2^p\right]
\end{equation}
with $\kappa =0.5$ GeV$^2/(Q^2+0.5$ GeV$^2$).  In Fig.~6 we show the 
two contributions in square brackets separately taking $\kappa=0.1$
which corresponds to choosing $Q^2=4.5$ GeV$^2$, i.e.\ $q=1$ as an
average of our considered fixed $q$-interval in (17).  The contribution
from the slope (first derivative) term $\dot{F}_2^p$ is indeed 
strongly suppressed and the curvature $\ddot F_2^p$ is the dominant
contribution in (18) in the small-$x$ region in NLO as well as in 
NNLO.  Since the suppression depends of course on the chosen value
for $Q^2$ in $\kappa$ we show in Table 2 the separate contributions
in square brackets in (18) calculated for three typical values of 
$Q^2$ in (17) at a fixed value of $x=10^{-4}$ in NLO and NNLO.  Even
at $Q^2=1.5$ GeV$^2$  $\ddot{F}_2^p$ dominates over $\kappa\dot{F}_2^p$
and therefore (18) represents a rather clean test of the curvature
of a structure function.

\section{Conclusions}

Perturbative NLO and NNLO QCD evolutions of parton distributions
in the (very) small-$x$ region are fully compatible with all 
recent high-statistics measurements of the $Q^2$-dependence of
$F_2^p(x,Q^2)$ in that region.  The results are perturbatively
stable and, furthermore, are rather insensitive to the 
factorization scheme chosen ($\overline{\rm MS}$ or DIS).
Therefore additional model assumptions concerning further
resummations of subleading small-$x$ logarithms are not required.
A characteristic feature of perturbative QCD evolutions is a
{\em{positive}} curvature $a_2(x)$ which {\em{increases}}
as $x$ decreases  (cf.~Fig.~5). This rather unique and perturbatively
stable prediction plays a decisive role in probing the range of
validity of perturbative QCD evolutions.  Although present data
are indicative for such a behavior, they are statistically 
insignificant for $x<10^{-4}$.  Future analyses of present
precision measurements in the very small-$x$ region should provide
a sensitive test of the range of validity of perturbative QCD and
further information concerning the detailed shapes of the gluon
and sea distributions as well. 
\vspace{0.4cm}

\noindent{\underline{\bf Acknowledgements}}

\noindent This work has been supported in part
by the   Bundesministerium f\"ur Bildung und Forschung,
Berlin/Bonn.

\newpage
%
%
\setlength{\oddsidemargin}{-0.5cm}
\begin{table}[th]
\normalsize
\renewcommand{\arraystretch}{1.8} 
\parbox{17cm}
{\caption[Table 1]{Parameter values of the NLO and NNLO QCD fits
with the parameters of the input distributions referring to (12)
and (13). Here $\chi^2$ was evaluated by adding in quadrature the
statistical and systematic errors.\newline}}
\vskip 15pt
\scriptsize
\centering 
\begin{tabular}{|c||c|c|c|c||c|c|c|c||c|c|c|c|}
\hline
& \multicolumn{4}{|c||}{NNLO($\overline{\rm MS}$)}  & 
\multicolumn{4}{|c||}{NLO($\overline{\rm MS}$)} &  
\multicolumn{4}{|c|}{NLO(DIS)}\\
\hline
& $u_v$ & $d_v$ & $\bar{q}$ & $g$ &
  $u_v$ & $d_v$ & $\bar{q}$ & $g$ &
  $u_v$ & $d_v$ & $\bar{q}$ & $g$\\
\hline
N & 0.2503 & 3.6204 & 0.1196 & 2.1961 &
    0.4302 & 0.3959 & 0.0546 & 2.3780 &
    0.6885 & 0.4476 & 0.0702 & 2.3445\\

a & 0.2518 & 0.9249 & -0.1490 & -0.0121 &
    0.2859 & 0.5375 & -0.2178 & -0.0121 &
    0.3319 & 0.5215 & -0.1960 & -0.0121\\ 

b & 3.6287 & 6.7111 & 3.7281 & 6.5144 &
    3.5503 & 5.7967 & 3.3107 & 5.6392 &
    2.6511 & 2.290 & 5.5480 & 6.8581\\

c & 4.7636 & 6.7231 & 0.6210 & 2.0917 &
    1.1120 & 22.495 & 5.3095 & 0.8792 &
    -1.6163 & 10.398 & 3.7277 & 1.8732\\

d & 24.180 & -24.238 & -1.1350 & -3.0894 &
    15.611 & -52.702 & -5.9049 & -1.7714 &
    15.197 & -16.466 & -4.7067 & -2.4302\\

e & 9.0492 & 30.106 & --- & --- &
    4.2409 & 69.763 & --- & --- &
    -7.6056 & 5.6364 & --- & ---\\
\hline
$\chi^2/{\rm dof}$ & \multicolumn{4}{|c||}{0.989} & 
                     \multicolumn{4}{|c||}{0.993} & 
                     \multicolumn{4}{|c|}{0.992}\\
\hline
$\alpha_s(M_Z^2)$ & \multicolumn{4}{|c||}{0.112} & 
                      \multicolumn{4}{|c||}{0.114} & 
                      \multicolumn{4}{|c|}{0.114}\\
\hline
\end{tabular}
\end{table}

\newpage
%
\setlength{\oddsidemargin}{-0.5cm}
\begin{table}[th]
\normalsize
\renewcommand{\arraystretch}{1.5} 
\parbox{17cm}
{\caption[Table 2]{The separate slope 
$\dot{F}_2^p\equiv \partial F_2^p/\partial \ln Q^2$ and curvature 
$\ddot{F}_2^p$ contributions to (18) in 
the $\overline{\rm MS}$ factorization scheme at $x=10^{-4}$ and
for the fixed $q$-interval in (17) with $\kappa = 0.5$ GeV$^2/
(Q^2+0.5$ GeV$^2$). The results are shown for three representative
values of $Q^2$ of this interval.  Notice that similarly to (15)
we used $F_2^p(x,Q^2) = A_0(x)+A_1(x)\ln Q^2+A_2(x)\ln^2Q^2$,
i.e.\ $\dot{F}_2^p=A_1+2A_2\ln Q^2$ and $\ddot{F}_2^p=2A_2$.
\newline}}
\vskip 15pt
\centering 
\begin{tabular}{|c||c|c|c||c|c|c|}
\hline
& \multicolumn{3}{c||}{NLO} & \multicolumn{3}{c|}{NNLO}\\

\hline
$Q^2/$GeV$^2$ & $\dot{F}_2$ & $\kappa\dot{F}_2 $ & $\ddot{F}_2$ &
              $\dot{F}_2$ & $\kappa\dot{F}_2 $ & $\ddot{F}_2$\\
\hline
1.5 & 0.3530 & 0.0883 & 0.1479 &
      0.3732 & 0.0933 & 0.1204\\

6   & 0.5580 & 0.0429 & 0.1479 &
      0.5401 & 0.0415 & 0.1204\\

12  & 0.6605 & 0.0264 & 0.1479 &
      0.6235 & 0.0249 & 0.1204\\
\hline
\end{tabular}
\end{table}
\newpage

\clearpage
\begin{figure}
\epsfig{file=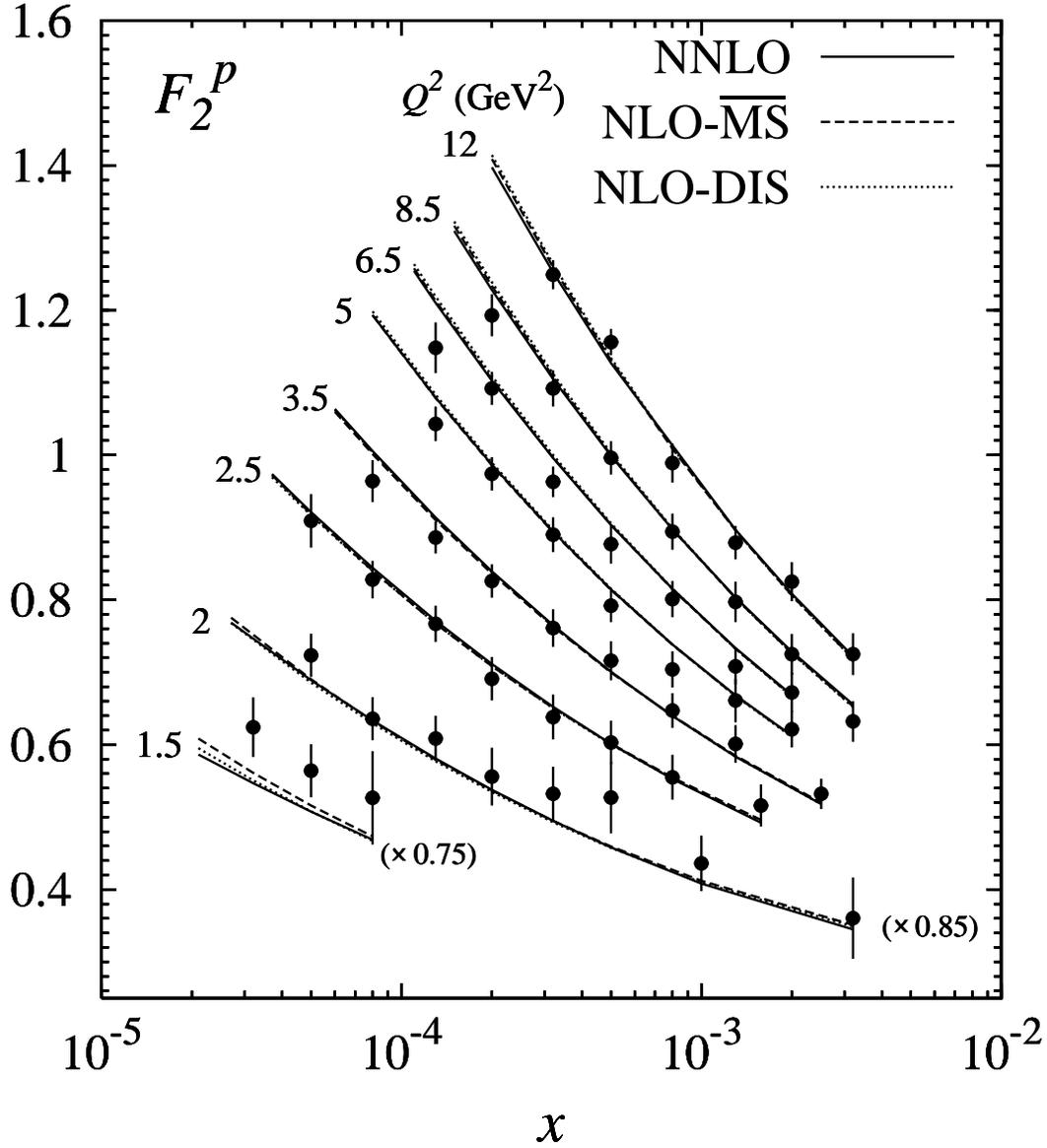, width=\textwidth}
\caption{Comparison of our various perturbative fits with the H1
data \cite{ref14} at very small-$x$.  Our 3-loop NNLO results 
always refer to the $\overline{\rm MS}$ factorization scheme.
To ease the graphical representation, the results and data for
the lowest two bins in $Q^2$ have been multiplied by the numbers
as indicated.}
\end{figure}

\clearpage
\begin{figure}
\epsfig{file=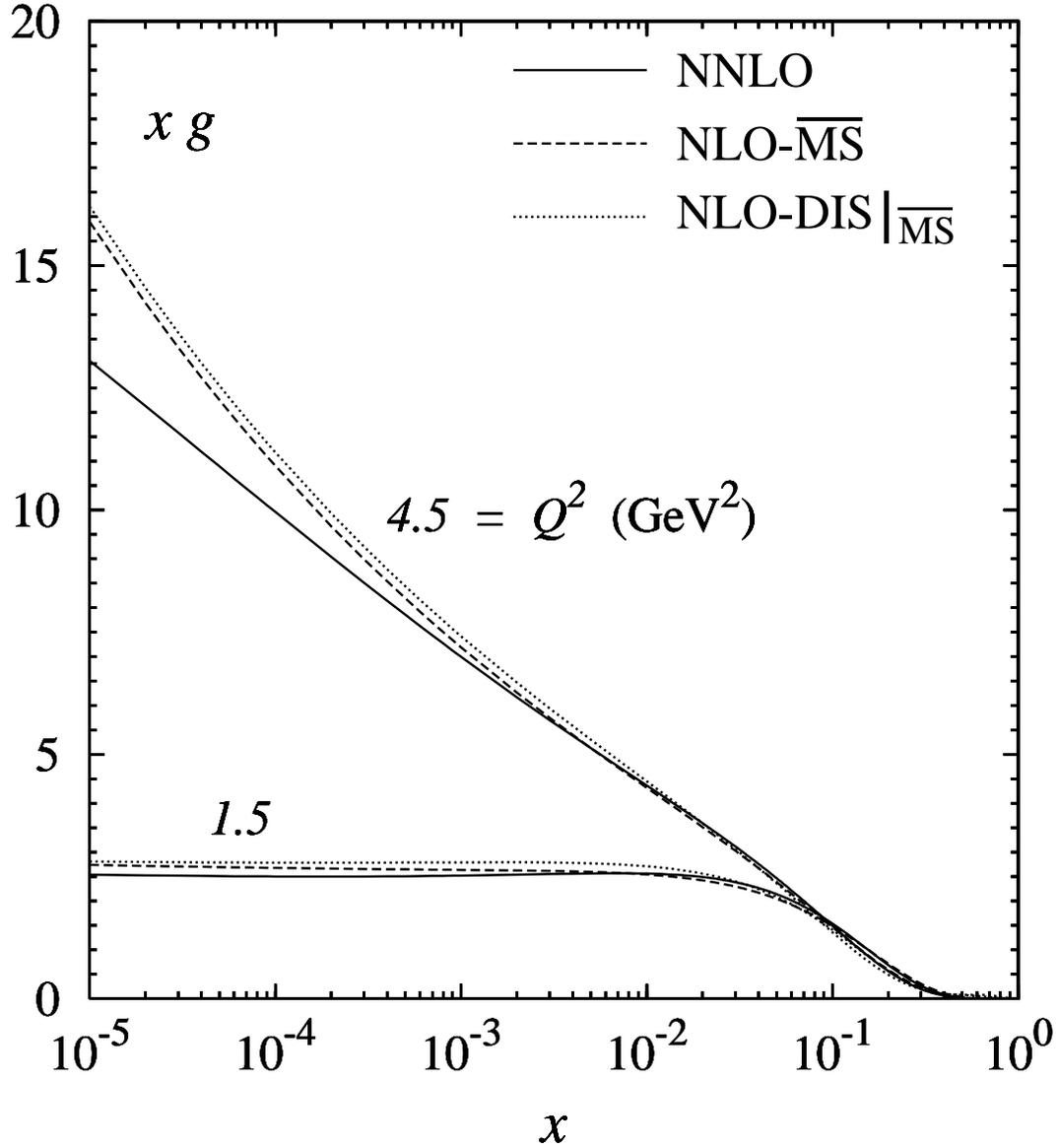, width=\textwidth}
\caption{The gluon distributions at the input scale $Q_0^2=1.5$ 
GeV$^2$, corresponding to (13) with the parameters given in Table 1,
and at $Q^2=4.5$ GeV$^2$.  For a consistent comparison with the
NNLO and NLO results in the $\overline{\rm MS}$ factorization
scheme, we have transformed our NLO-DIS results to the 
$\overline{\rm MS}$ scheme using (9) which are denoted by
NLO-DIS$|_{\overline{\rm MS}}\,$.}
\end{figure}

\clearpage
\begin{figure}
\epsfig{file=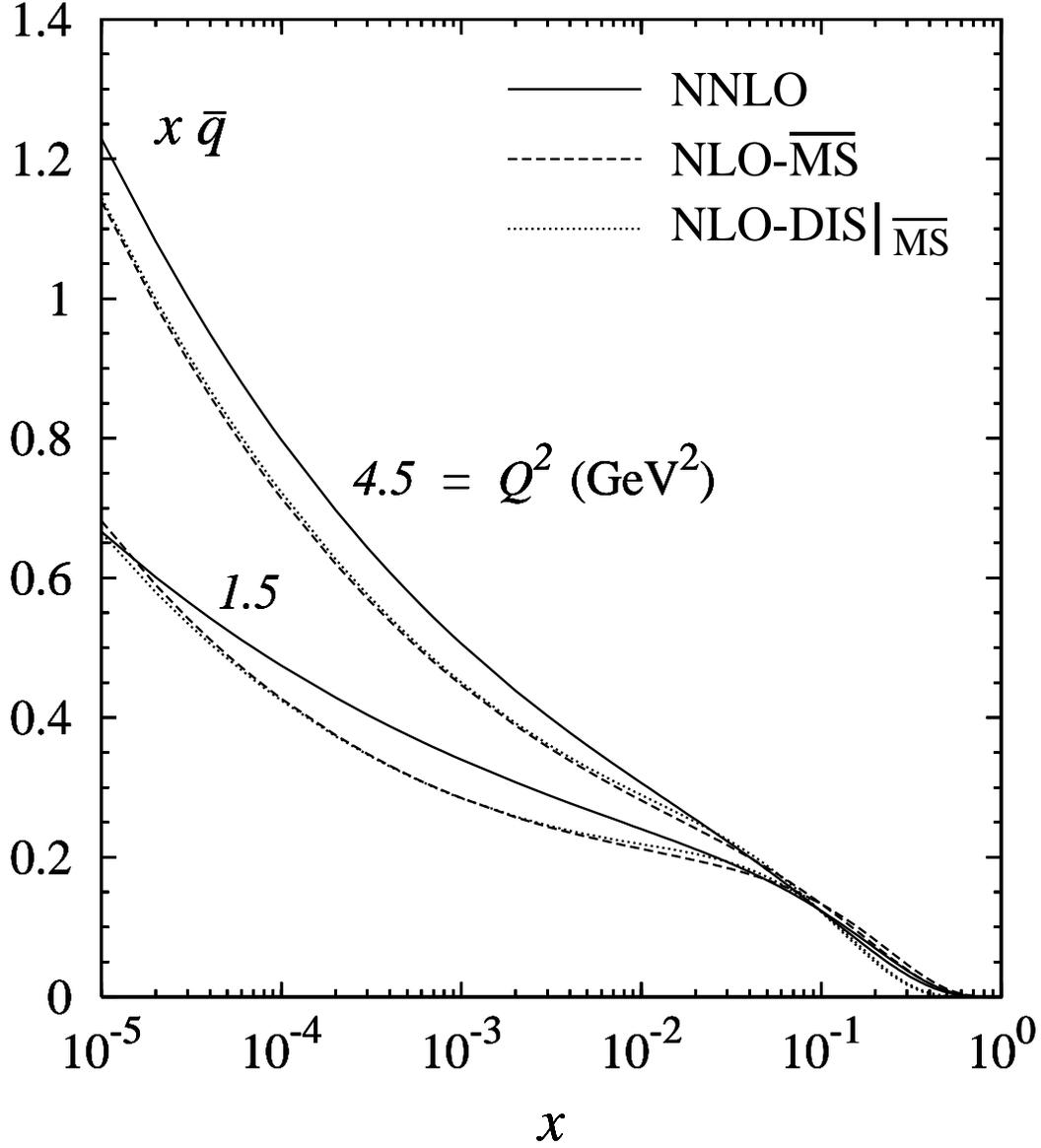, width=\textwidth}
\caption{As in Fig.~2 but for the sea distribution $x\bar{q}(x,Q^2)$
where $\bar{q}\equiv\bar{u}=\bar{d}$.  The NLO-DIS results have
been transformed to the $\overline{\rm MS}$ factorization scheme
using (8) which are denoted by NLO-DIS$|_{\overline{\rm MS}}\,$.}
\end{figure}

\clearpage
\begin{figure}
\epsfig{file=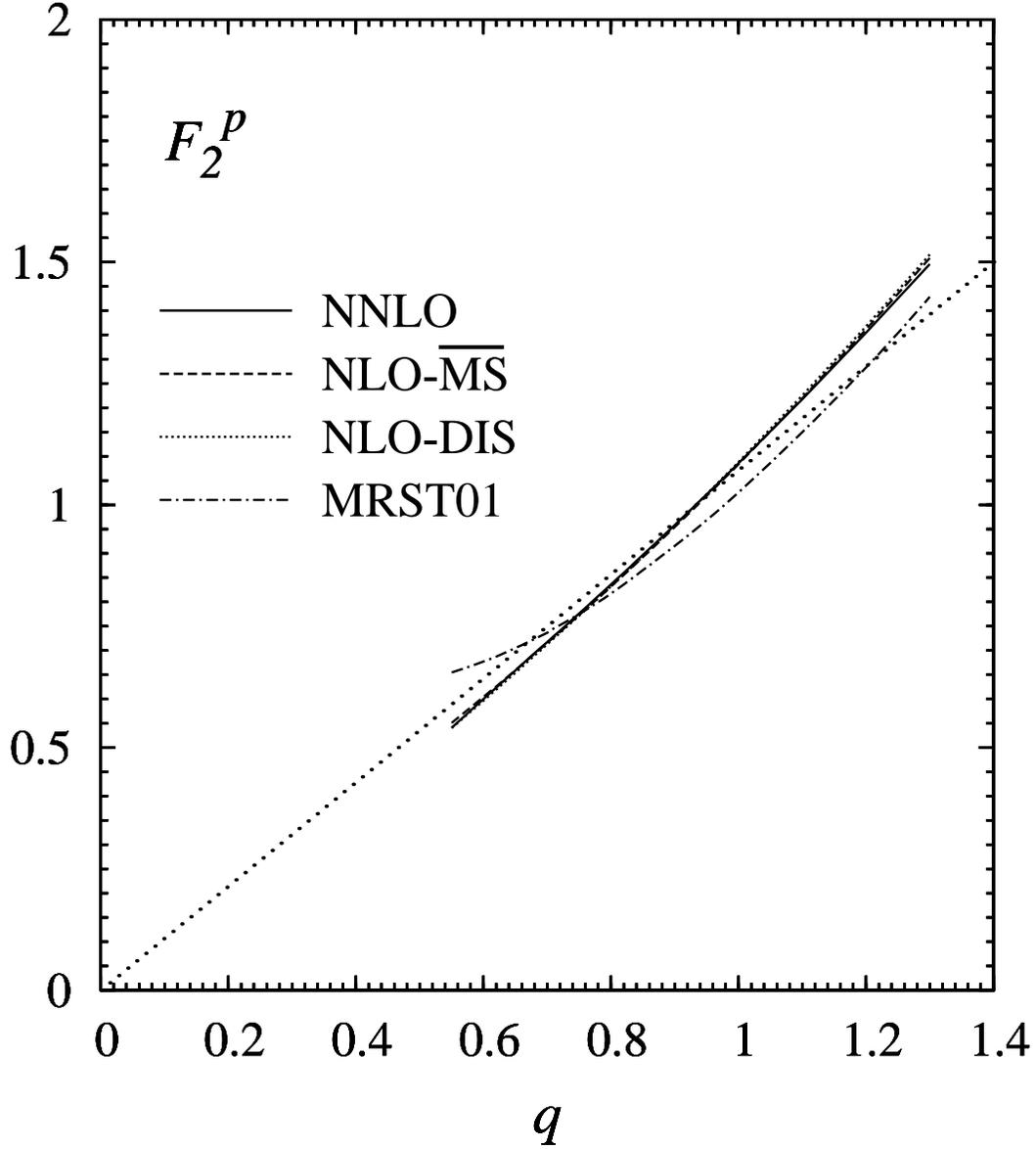,width=\textwidth}
\caption{Predictions for $F_2^p(x,Q^2)$ at $x=10^{-4}$ plotted
versus $q$ defined in (14).  For comparison the global fit NLO result
of MRST01 \cite{ref20} is shown as well.  The global CTEQ6M NLO fit
\cite{ref21} is very similar to our NLO and NNLO results as can
be deduced from \cite{ref2}, and the same holds true for the H1
fit \cite{ref15}.  Most small-$x$ data lie along the straight
dotted line \cite{ref1}.} 
\end{figure}

\begin{figure}
\begin{center}
\vspace*{-4cm}
\epsfig{file=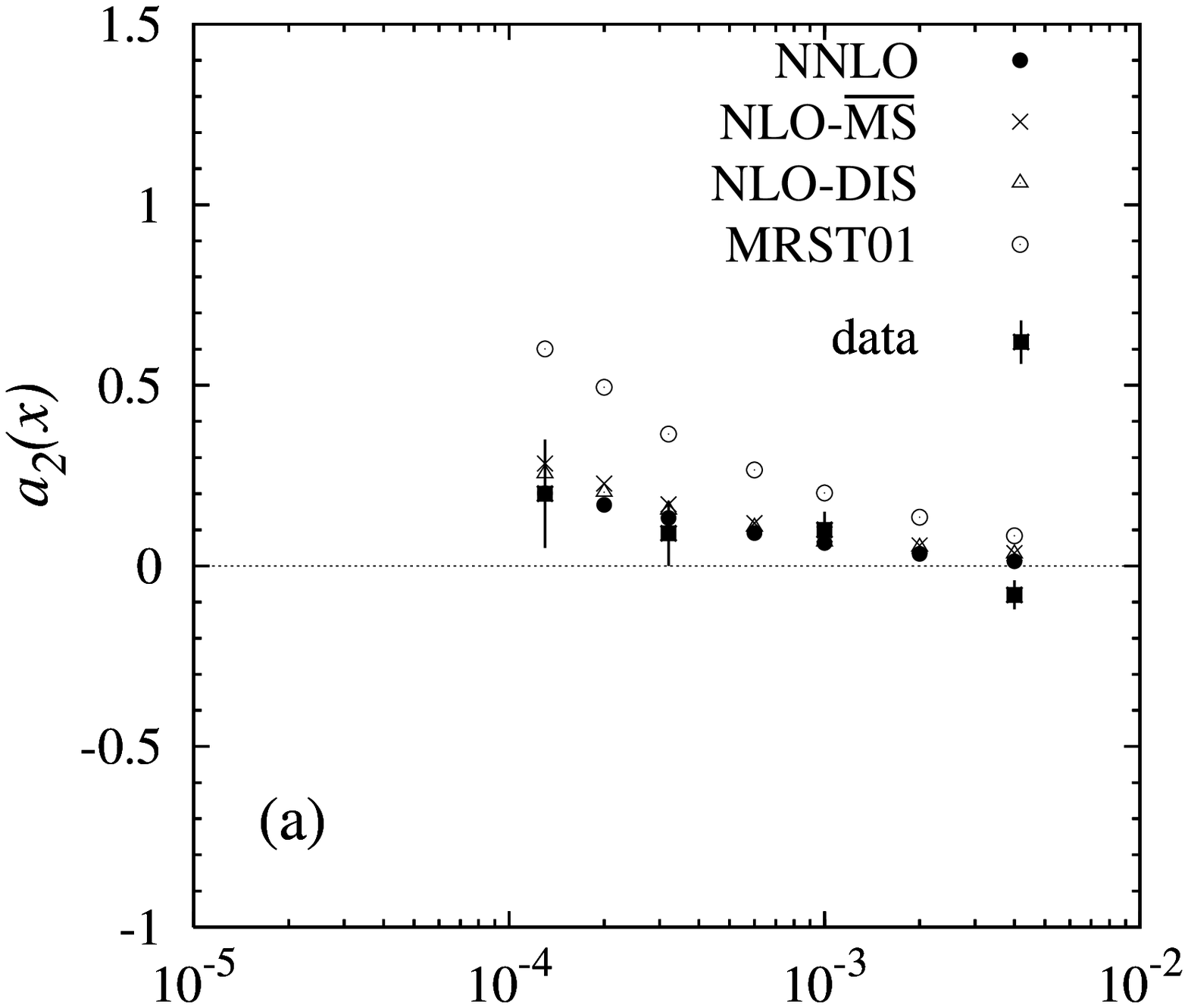, width = 13 cm,height = 10 cm}
\epsfig{file=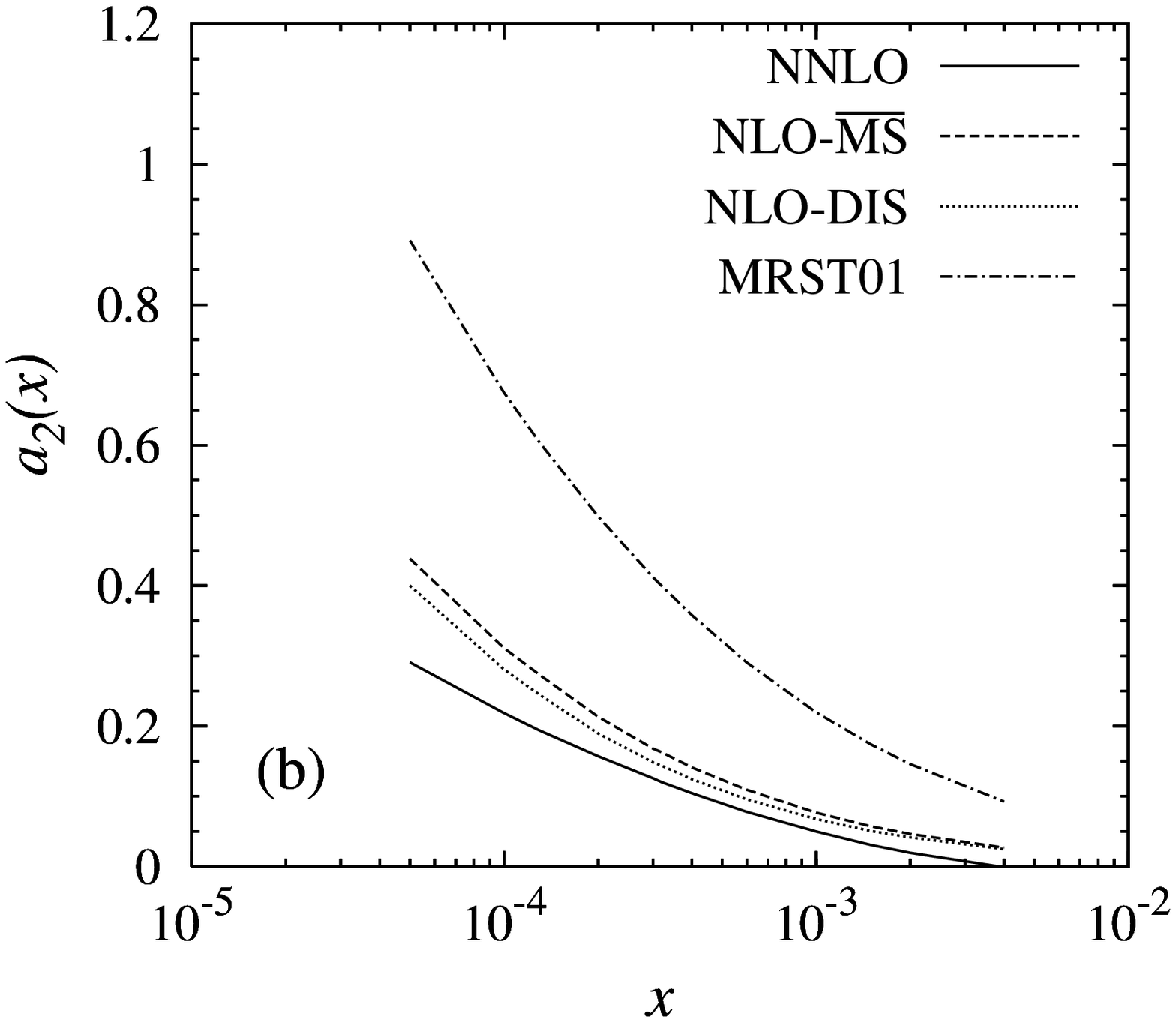, width = 13 cm,height = 10 cm}
\end{center}
\caption{The curvature $a_2(x)$ as defined in (15) for (a) the
variable $q$-intervals in (16) and (b) the fixed $q$-interval in
(17).  Also shown are the corresponding MRST01 NLO 
\mbox{results \cite{ref20}.}
The data in (a) are taken from \cite{ref1}.  The NNLO prediction
at the lowest $x$-value coincides with the data (full square).}  
\end{figure}


\clearpage
\begin{figure}
\epsfig{file=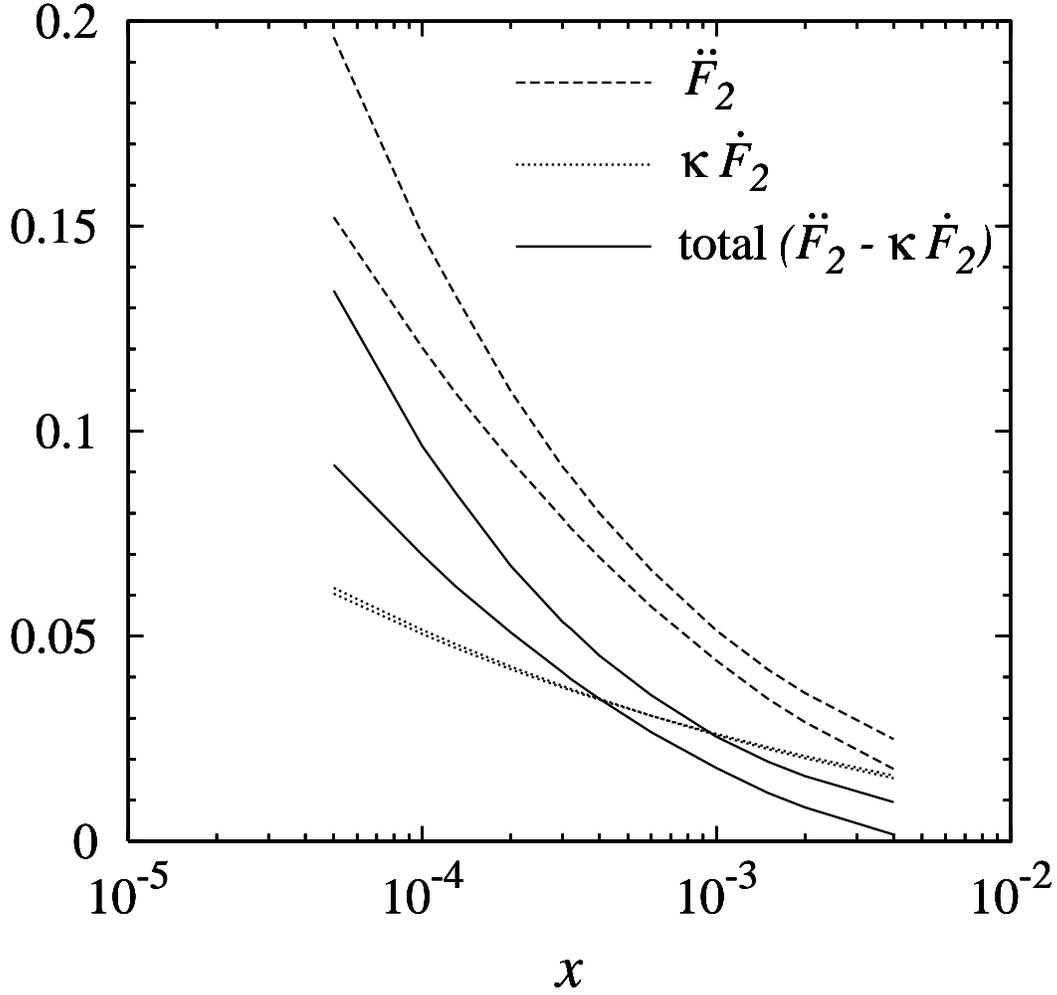,width=\textwidth}
\caption{The predicted slope 
$\dot{F}_2^p\equiv\partial F_2^p/\partial\ln Q^2$
and curvature $\ddot{F}_2^p$ appearing in (18) for the fixed
$q$-interval in (17), with the suppression factor $\kappa=0.1$
corresponding to an average $Q^2=4.5$ GeV$^2$ ($q=1$).  At smallest
values of $x$, the individual upper curves always refer to 
NLO($\overline{\rm MS}$) and the lower ones to 
NNLO($\overline{\rm MS}$).}
\end{figure}
\end{document}